\documentclass[a4paper,12pt,titlepage]{article}
\usepackage{amsmath}
\usepackage{amsfonts}
\usepackage{graphicx}
\usepackage{scrextend}
\usepackage{authblk}
\usepackage{algorithmic}
\usepackage{algorithm}
\usepackage{amsthm}
\usepackage[numbers,comma,sort&compress]{natbib}
\usepackage{float}
\usepackage{lineno}
\usepackage{multirow}
\usepackage[utf8]{inputenc}

\renewcommand{\phi}{\varphi}
\renewcommand{\epsilon}{\varepsilon}

\newcommand{\mathscr}{\mathcal} 

\def\squareforqed{\hbox{\rlap{$\sqcap$}$\sqcup$}}
\def\qed{\ifmmode\squareforqed\else{\unskip\nobreak\hfil
\penalty50\hskip1em\null\nobreak\hfil\squareforqed
\parfillskip=0pt\finalhyphendemerits=0\endgraf}\fi}

\newtheorem{control-problem}{Control Problem}

\newcommand{\ignore}[1]{}

\begin{document}

\renewcommand{\thefootnote}{\fnsymbol{footnote}}
\title{\vspace{-10ex}{Multimodal Cross-registration and Quantification of Metric Distortions in Whole Brain Histology of Marmoset using Diffeomorphic Mappings}}
\author[1,2]{Brian C. Lee \thanks{Corresponding author: blee105@jhu.edu (BCL)}}
\author[3]{Meng K. Lin}
\author[4]{Yan Fu}
\author[3]{Junichi Hata}
\author[1,2]{Michael I. Miller}
\author[5]{Partha P. Mitra}
\affil[1]{Center for Imaging Science, Johns Hopkins University, Baltimore, MD, USA}
\affil[2]{Department of Biomedical Engineering, Johns Hopkins University, Baltimore, MD, USA}
\affil[3]{RIKEN Center for Brain Science, Wako, Japan}
\affil[4]{Shanghai Jiaotong University, Shanghai, China}
\affil[5]{Cold Spring Harbor Laboratory, Cold Spring Harbor, NY, USA}
\date{\vspace{-5ex}}
\maketitle
\renewcommand{\thefootnote}{\arabic{footnote}}

\begin{abstract}
Whole brain neuroanatomy using tera-voxel light-microscopic data sets is of much current interest. A fundamental problem in this field, is the mapping of individual brain data sets to a reference space. Previous work has not rigorously quantified the distortions in brain geometry from in-vivo to ex-vivo brains due to the tissue processing. Further, existing approaches focus on registering uni-modal volumetric data; however, given the increasing interest in the marmoset, a primate model for neuroscience research, it is necessary to cross-register multi-modal data sets including MRIs and multiple histological series that can help address individual variations in brain architecture. These challenges require new algorithmic tools. 

Here we present a computational approach for same-subject multimodal MRI guided reconstruction of a series of consecutive histological sections, jointly with diffeomorphic mapping to a reference atlas. 
We quantify the scale change during the different stages of histological processing of the brains using the determinant of the Jacobian of the diffeomorphic transformations involved. There are two major steps in the histology process with associated scale distortions (a) brain perfusion (b) histological sectioning and reassembly. By mapping the final image stacks to the ex-vivo post fixation MRI, we show that tape-transfer assisted histological sections can be re-assembled accurately into 3D volumes with a local scale change of $2.0 \pm 0.4\%$ per axis dimension. In contrast, the perfusion/fixation step, as assessed by mapping the in-vivo MRIs to the ex-vivo post fixation MRIs, shows a significantly larger median absolute scale change of $6.9 \pm 2.1\%$ per axis dimension.  This is the first systematic quantification of the local metric distortions associated with whole-brain histological processing, and we expect that the results will generalize to other species. It will be important to take these local scale changes into account when computing properties such as local cell and process densities at the voxel level in creating reference brain maps. 
\end{abstract}

\section{INTRODUCTION}
The common marmoset (\textit{Callithrix jacchus}) is an increasingly important model species for both scientific and preclinical neuroscience research. As compared with other non-human primates, the marmoset brain has unique advantages for brain mapping including a compact brain size, a lissencephalic cortex, and the availability of transgenic lines \cite{Sasaki-2009}, \cite{Kishi-2014}. However, mapping the complete neural circuitry of the marmoset brain is a time consuming and complex process that requires a coordinated effort involving multidisciplinary scientific knowledge and collaboration. One important task in any circuit mapping project is to map the whole-brain neuroanatomical data to a reference atlas. This is more challenging for primate species such as the marmoset when compared with commonly used animal models such as the mouse, due to higher individual variability in brain geometry and the increased complexity of the brain architecture of the primate brain. To address these issues it is necessary to obtain multimodal data sets showing different kinds of contrast, including in-vivo and ex-vivo magnetic resonance imaging (MRI) volumes in combination with whole-brain LM images of a series of histologically labelled sections, cf. Nissl stains, myelin stains and the use of fluorescent probes.

In this paper, we focus on the computational problem mapping of serially-sectioned whole-brain image datasets of different modalities and scales into the same coordinate space.
In particular, we describe the registration procedure on a specific experimental methodology for obtaining brain-wide connectivity information using neurohistological preparations, where thinly sliced brain sections are systematically subjected to different histological procedures in order to obtain a broad spectrum of information about cytoarchitectonic structure. The sections were obtained using a high-throughput neurohistological pipeline that has been previously described \cite{Lin-2019-elife}, along with in-vivo and ex-vivo MRI volumes. Our results demonstrate that the local metric distortions of the brain after histological preparations are relatively small with the exception of the perfusion-fixation step. Our results provide a systematic characterization of the metric distortions associated with the perfusion-fixation step, which can be generalized to the brains of other species as well. 
Although much of the presented data is produced by histology, we include as a subset methods for 3D image data generated by techniques (eg MRI) that do not require section-registration. The analysis of in-vivo to ex-vivo MRI mappings described in this paper do not require the section registration step.  

{\it Prior work on artifacts associated with reassembling 2D sections} Light-microscopic histological image section reconstruction guided by reference imagery has been previously explored by several groups. The problem of accumulated long range distortions in unguided reconstruction was studied by Malandain \cite{Malandain2004} and one of the earlier proposed guided-reconstruction solution was Block-matching i.e. using block-face photographs during histology to align sections \cite{Dauguet2007}. Feature extraction and manually defined landmarks have also been applied to guide reconstruction \cite{Streicher1997}. More recent methods have used same-subject MRI to guide reconstruction of human \cite{Adler2014} and mouse \cite{Yang-2012} brain structures. None of these studies however provide the metric quantification reported in this paper. We believe our methods constitute a significant advance that will be of particular importance for quantitative studies, in particular for large volume brains such as that of non-human primates and humans. 

{\it Prior work on diffeomorphic methods for registering brain volumes} Diffeomorphic mapping of brain volumes has been a central focus in the field of Computational Anatomy
\cite{GrenanderMiller1998,Dupuis1998,Miller-Younes-2001,Toga2001,Miller2002,Thompson2002,Miller2004-growth,beg2004computational,joshi2004unbiased,Ashburner2007,GrenanderMiller2007,Durrleman2008b,Ashburner2009,younes2009evolutions,Younes2010,Pennec2011,adams2013computational}.
Mapping methods initially followed the small deformation and elasticity methods of Bajcsy and others \cite{Bajcsy-Lieberson-Reivich-1983,BajcsyKovacic1989,AmitGrenander1989,Gee93elastic,Miller-PNAS-1993,Rabbitt-Weiss-1995}.
Subsequently Christensen et al \cite{Christensen1996}
introduced large deformation flows for topology preservation in dense volume matching \cite{Christensen97volumetrictransformation}.
Since these early inceptions many methods have been developed based on both landmark and triangulated surface based spline deformations \cite{Bookstein:1989:PWT:66131.66134,Bookstein1991thin,Bookstein:1996} as well as large deformation methods \cite{JoshiMiller2000,CamionYounes2001,glaunes2004diffeomorphic, Vaillant05surfacematching,GlaunesQiu2008}.
For dense images (i.e. 3D voxelized image volumes) with multiple modalities and tensor fields such as Diffusion track imaging (DTI), these methods were further developed and form the basis of the multiple contrast, multi-scale algorithmic framework that is well-described in the literature, but note that much of this work has focussed on MRI volumes as opposed to the LM based teravoxel volumes of interest to the present study \cite{avants2004geodesic,Beg2005,beg2007symmetric,vercauteren2008symmetric,avants2008symmetric,CeritogluMultiModal2009,ceritoglu2010large,Risser2011,Sommer2011,twardceritoglu2011,DuYounesQiu2011,VadakkumpadanTrayanova2012,Tang-Miller-Mori-2012,du2012diffeomorphic,du2014diffeomorphic,risser2013piecewise,Tward2013,khan2013multistructure}.

{\it Prior work on histology-associated deformations} Deformation of brain tissue caused by histological procedures is well known and has been reported in the previous literature. The chemical composition of the fixation solution and duration of exposure have been previously shown to cause significant tissue shrinkage \cite{Mouritzen-1979, Leibnitz-1971, Quester-1997}. More recently, histological distortions have been quantified with imaging techniques like MRI or computed tomography using variables such as total brain volume and the distance between hand-selected landmarks, in comparisons before and after mouse brain histology \cite{Wehrl-2015}. Others have assessed dense local deformative effects of extraction and fixation by examining the strain resulting from a non-rigid displacement field \cite{Schulz-2011}. Comparison of the total tissue area of imaged histological sections with block-face images can quantify the global in-plane shrinkage caused by sectioning, and neuronal density in the cutting axis has been used to quantify the non-uniformity of shrinkage in that direction \cite{Carlo-2011}.

{\it What we do:} Here, we propose an automated computational pipeline for same-subject MRI reference guided reconstruction of multi-modal histological image stacks of the marmoset brain followed by registration to a labeled atlas. 
We build upon our recent work incorporating Sobolev priors to solve the histological stacking problem of Nissl stacks onto high resolution atlases at the light microscope scale \cite{Lee-Mitra-Miller-PLOS-2018}. However, this previous work only employed a standard reference atlas, and did not have access to same-brain MRIs. In the present paper, our reconstruction process is informed by the ex-vivo MRI of the brain prior to histology as well as an image intensity smoothness prior. To perform robust atlas-mapping, we have used a variant of the multi-channel large deformation diffeomorphic metric mapping (LDDMM) algorithm which applies voxel-level weights to the image similarity metric to account for contrast changes, as well as damage or noise in histological sections. 

We apply our new methodology to a dataset of marmoset brains obtained as part of the RIKEN Brain/MINDS project and demonstrate solutions to the multi-modal stack reconstruction problem as well as robust atlas mapping results across four modalities. 
Importantly, we are able to quantify the 3D tissue distortion caused by two major parts of the histology procedure -- the ``sectioning'' process (cryoprotection, freezing, sectioning) captured by the ex-vivo MRI to histology mapping and ``preparatory'' process (injection, incubation, perfusion, fixation) captured by the in-vivo MRI to ex-vivo MRI mapping. The former histological process causes only a $\sim 2\%$ scale change (median absolute local scale change), whereas the perfusion-fixation step causes a much larger scale change of $\sim 7\%$.

\section{RESULTS}

\begin{figure}[H]
\includegraphics[width=0.9\textwidth]{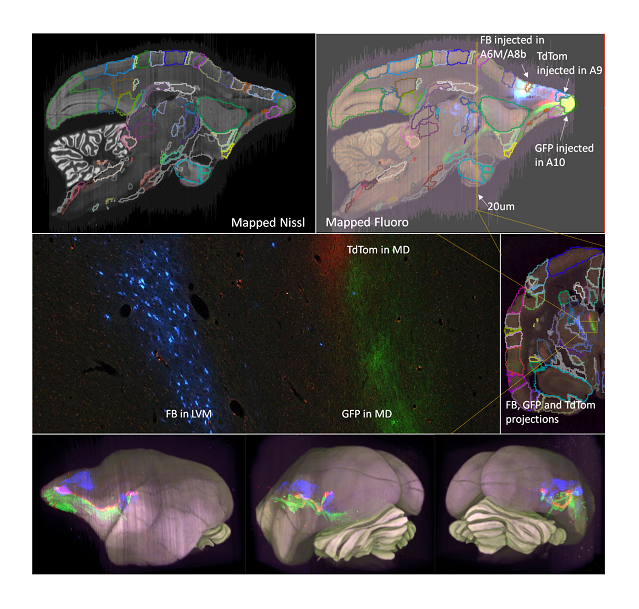}
\caption{Top two rows: Nissl and fluorescent reconstructed segmented volumes. NB: the sections are originally cut in the coronal section; a sagittal virtual cut of the 3D reconstructed brain is shown. The upsampled reconstruction transforms are applied to the full resolution fluorescent tracer images where tract tracing can be performed. Here, the three injected tracers are labeled in the high resolution image (bottom left). Bottom row: 3D visualization of the Nissl stack reconstruction overlayed with the red, green, and blue tagged tracer paths detected from the registered fluorescence volumes.
\label{fig:segmentation}
\label{fig:3D_visualize}}
\end{figure}

Brain region parcellations were generated by the proposed joint optimization of an LDDMM-based atlas mapping procedure with stack reconstruction of a Nissl-stained histological volume. The histological restacking was guided by a same-subject ex-vivo MRI scan. These informed variational solutions were optimized for each of fifteen brains in the Brain/MINDS dataset and we observed accurate estimates of the reconstructed stacks consistent with simulations previously reported in \cite{Lee-Mitra-Miller-PLOS-2018}. Sample segmentation and process detection results are depicted in Figure \ref{fig:segmentation} for an individual marmoset brain. 
Segmentation of the fluorescent images was achieved using the same reconstruction framework that was applied to the original Nissl stack, by using the corrected Nissl stack as an exact shape prior with only rigid in-plane cross-registration connecting the two series. 
Figure \ref{fig:segmentation} shows examples of fluorescent image stacks being reconstructed by transferring the segmentation computed on the Nissl stack. The top two rows of Figure \ref{fig:segmentation} show the computed transforms applied to the full resolution fluorescence image stack. Major connections and fiber tracts originating from the fluoroscent tracer injections (Red: TRE3-tdTomato anterograde; Green:TRE3-Clover anterograde; Blue: Fast Blue retrograde) can be identified from the color-coded stains in 3D. The bottom row of
Figure \ref{fig:3D_visualize} shows 3D reconstructions of the tracings.

We examine the first fundamental form, or determinant of the metric tensor, of the diffeomorphic mapping as described in Eqn. \eqref{jacobian-matrix} (see Methods section 4.5). We do this to quantify the distortion caused by the histological process. Several preparatory processes occur in between the in-vivo MRI acquisition and the ex-vivo MRI acquisition -- we interpret the Jacobian determinant of the mapping between these coordinates spaces as the combined deformative effect of tracer injection, extraction, perfusion, and tissue fixation. We report the "percent scale change factor" which is computed as $|\det \partial_X \varphi|^{\frac{1}{3}}$, the cube root of the Jacobian determinant, and which represents the per-axis local scale change. Similarly, the freezing and sectioning processes occur between the ex-vivo MRI acquisition and the histological imaging. Under our informed histological reconstruction model of Section 4.2 we can interpret the scale change between these two coordinate spaces as the combined deformative effect of freezing/sectioning.

\begin{figure}[H]
	\begin{center}
\includegraphics[width=1.0\textwidth]{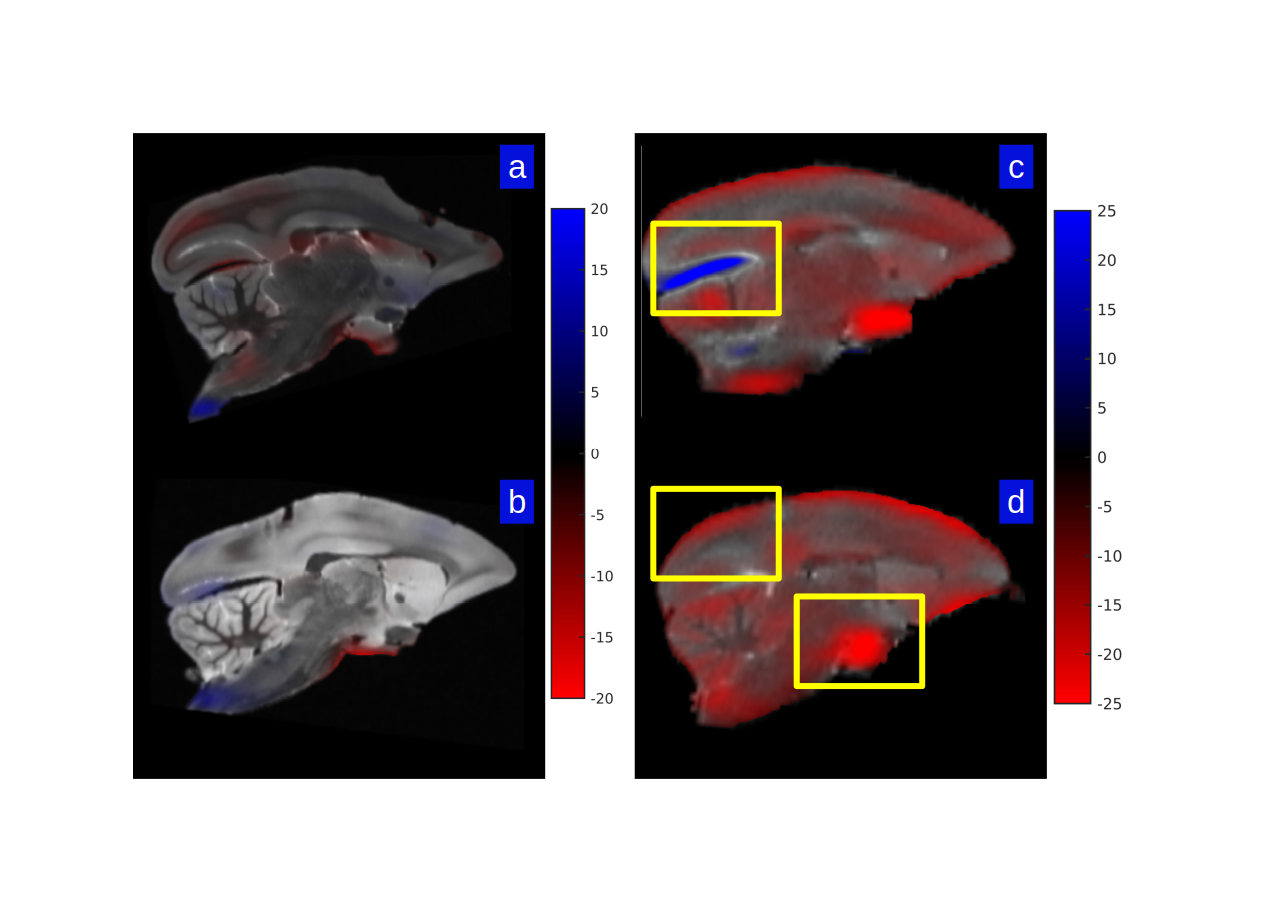}
\caption{Shown is the percent scale change away from the identity of the mapping as measured by the cube-root of Jacobian determinant; blue depicts shrinkage, red expansion. Panels A and B show two examples of ex-vivo MRI mapped to Nissl histological stack; panels C and D show the same for the
ex-vivo to in-vivo MRI.
Yellow boxes depict intense scale changes which are depicted via grid deformation shown in Figure \ref{fig:defgrid_jacobian_comparison}.
\label{fig:jacobian}}
\end{center}
\end{figure}

We illustrate the quantitative properties of the Jacobian matrix as a first-order description of the map. 
Four sample measurements (two of the sectioning process, A and B, and two of the preparatory process, C and D) from the dataset are shown in Figure \ref{fig:jacobian}. These demonstrate that there is minimal metric scale change away from the identity map for the ex-vivo MRI to sectioned histology maps. However, the measured metric scale was much higher for the in-vivo pre-preparatory MRI to ex-vivo post-preparatory MRI maps.
Panels A and B show the percentage metric change away from the identity of the cube-root of the Jacobian between ex-vivo MRI and the Nissl reconstructed brains.
Shown as a heat-map superimposed over the gray level images is the cubed root of the Jacobian determinant for the central sagittal section of each subject for each of two brains for mapping ex-vivo to histology stack section. As depicted by the color bar, the maximum value of blue represents 20 percent expansion in a dimension, with the red implying contraction.
Panels C and D show similar analyses for two brains corresponding to ex-vivo to in-vivo MRI maps, indicating several areas of significant contraction and expansion. 
Overall, we report the mean across 15 subjects of the median absolute percent scale change as 1.97 $\pm$ 0.38 \% for the sectioning process (ex-vivo MRI to reconstructed histology mapping) and 6.90 $\pm$ 2.08 \% for the fixation process (ex-vivo MRI to in-vivo MRI mapping).


\begin{figure}[H]
	\begin{center}
\includegraphics[width=0.8\textwidth]{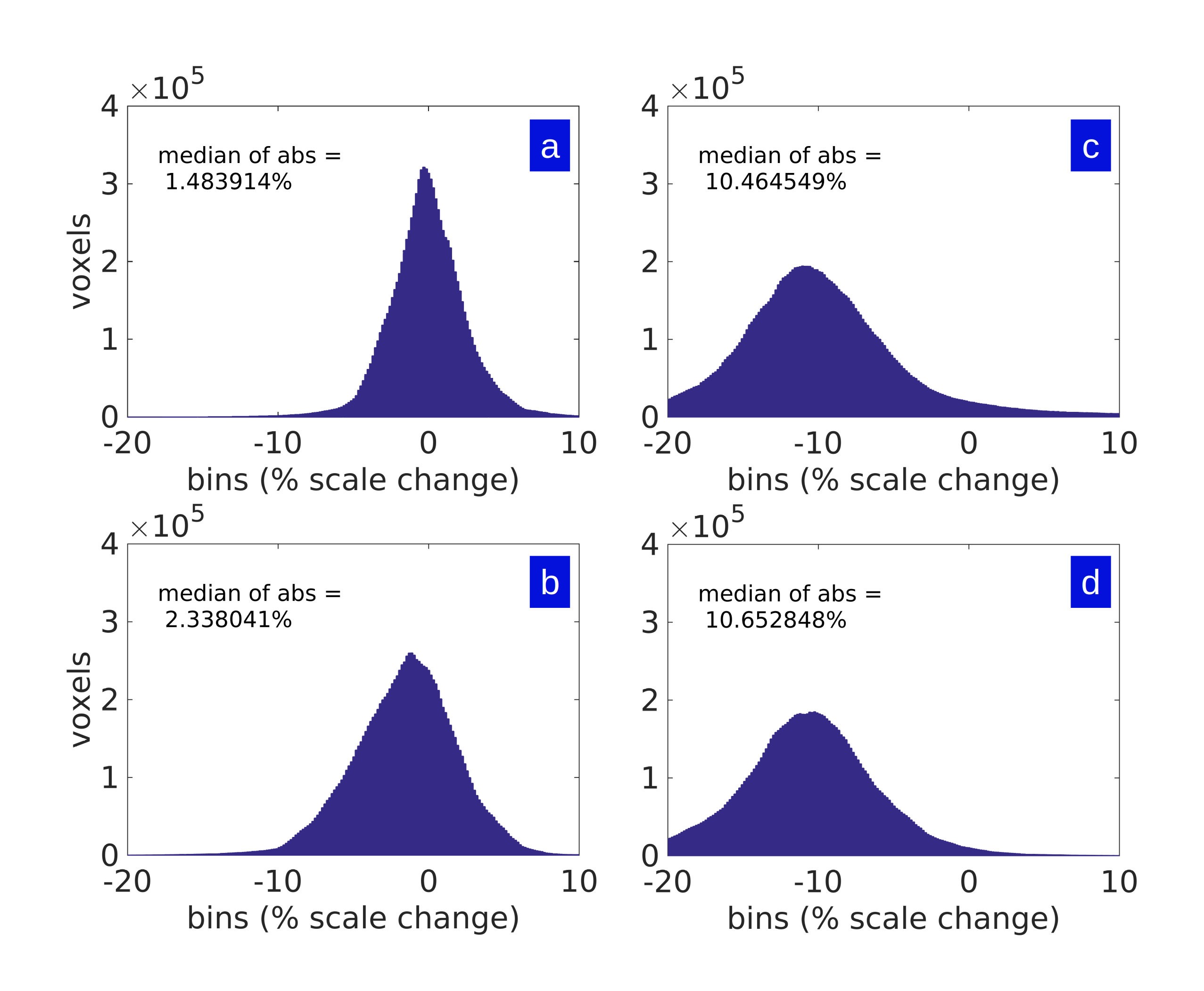}
\caption{Shown are histograms of the percent scale in single axis dimension as measured by cube-root of Jacobian determinant of maps shown in Figure \ref{fig:jacobian}. Panels A and B show the ex-vivo MRI mapped to Nissl histological stack; Panels C and D show the ex-vivo MRI to in-vivo MRI histograms.
\label{fig:jacobian-histograms-only}}
\end{center}
\end{figure}

Shown in Figure \ref{fig:jacobian-histograms-only} are histograms of the percent scale change factor between the ex-vivo to Nissl histological stack (A \& B) and the ex-vivo to in-vivo spaces (C \& D) for the same brains shown in Figure \ref{fig:jacobian}.

\begin{figure}[H]
	\begin{center}
\includegraphics[width=0.8\textwidth]{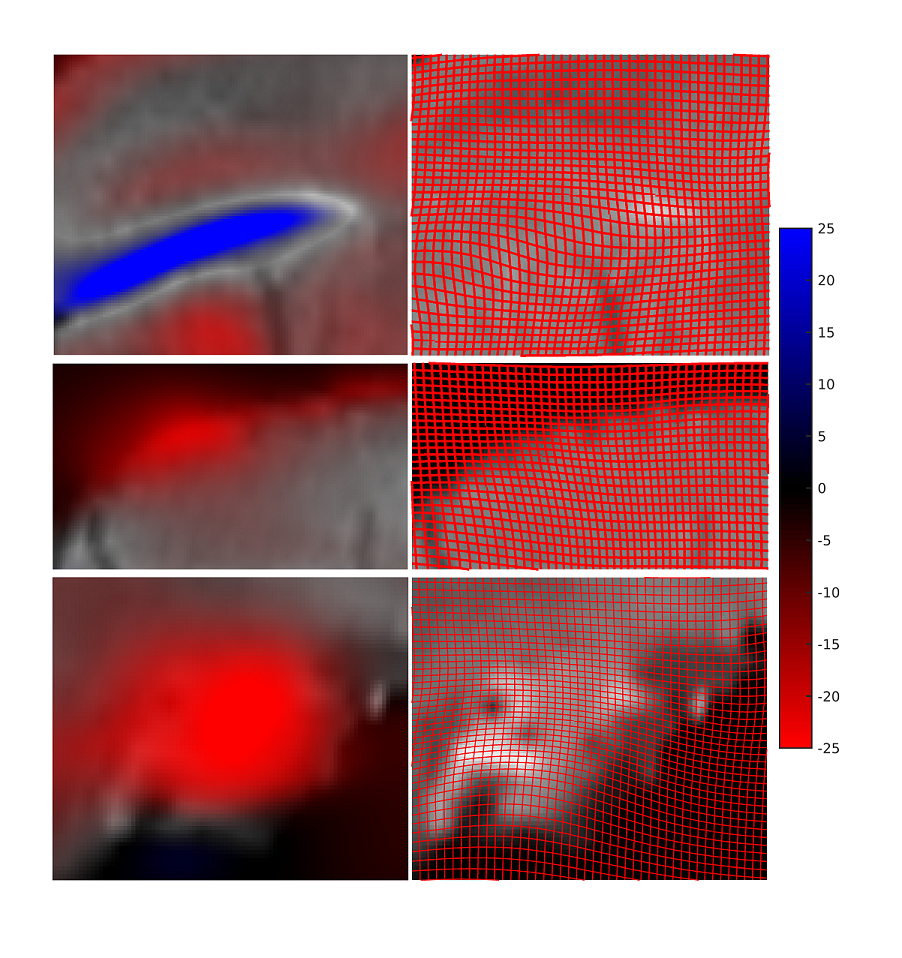}
\caption{Metric scale change associated with subvolume sections between in-vivo and ex-vivo MRIs. Maximum blue indicates 25 percent expansion, maximum red indicates 25 percent contraction as measured by the cubed-root of Jacobian determinant.
Right column shows the how a uniform square grid on the in-vivo brain deforms when mapping to the ex-vivo brain. This helps visualize the scale factor change from identity. Both grid expansion (top row blue) and grid contraction (bottom row red) is seen.
\label{fig:defgrid_jacobian_comparison}}
\end{center}
\end{figure}

As seen in Figures \ref{fig:jacobian} and \ref{fig:jacobian-histograms-only},
the change in measures between the ex-vivo post-preparatory and in-vivo MR-measured coordinate systems are large and mostly contractive in each dimension as measured by the cubed-root of Jacobian determinant, In comparison, the corresponding metric change due to the sectioning process are small, and are almost symmetric around zero (so that both shrinkage and expansion occurs in roughly equal proportions). The histograms of Figure \ref{fig:jacobian-histograms-only} demonstrate that the range of the ex-vivo post-preparatory to in-vivo MR maps are nearly three times in terms of median absolute scale change.
Figure \ref{fig:defgrid_jacobian_comparison} shows that scale change can be as much as 20-25 percent in a single axis dimension as measured by the cubed-root of Jacobian determinant.
Shown are sample in-plane deformations for magnified brain regions from Figure \ref{fig:jacobian}.
Notice that the intense blue indicates 25 percent expansion, and intense red indicates 25 percent contraction along a single axis dimension.

\begin{figure}[h]
\begin{center}
\vspace{-0.3cm}
\begin{tabular}{cc}
\includegraphics[width=1.0\textwidth]{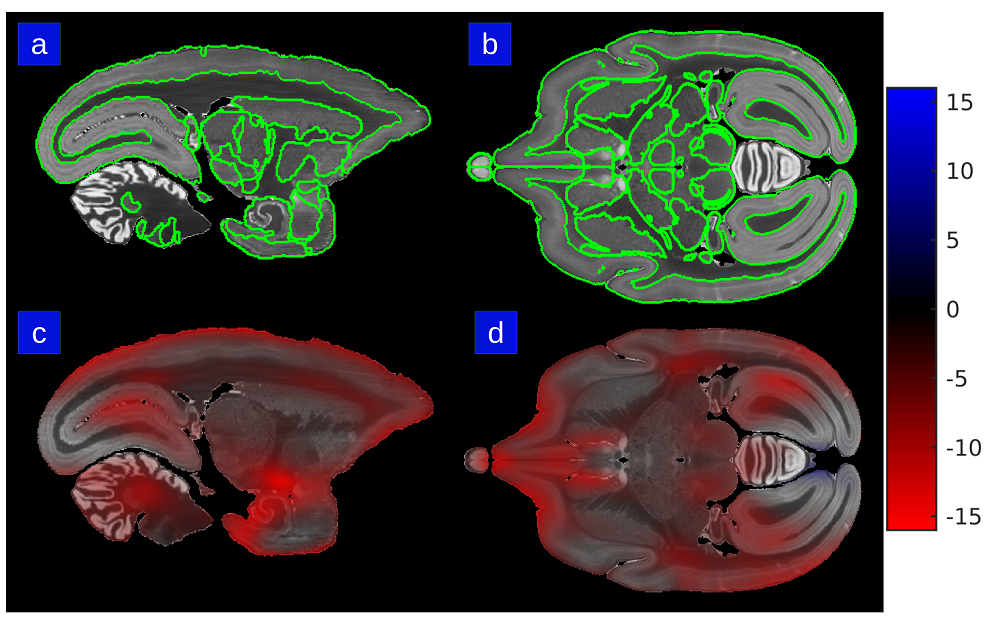}
\end{tabular}
\vspace{-0.2cm}
\caption{
Metric scale change of in-vivo to ex-vivo mapping in atlas coordinates
with heat map superimposed on olfactory bulb, cortical areas, septum, thalamus, epithalamus, pretectum, brainstem, hypothalamus, cerebellum, cranial nerve, entorhinal cortex.
Panels A \& B show section outlines of structures in the Paxinos/Hasikawa atlas in sagittal and transverse views. Panels C \& D show the metric scale change for the same sections. The color bar depicts maximum red of percent scale contraction in a single axis dimension.
\label{fig:jacobian-mean-exvivo-invivo}
}
\end{center}
\end{figure}

A closer examination of the brain-wide distribution of changes in tissue volume due to the preparatory process between the in-vivo and the ex-vivo MR images is shown in Figure \ref{fig:jacobian-mean-exvivo-invivo}. The heat map of the percent scale change factor of one mapping from in-vivo to ex-vivo MRI is shown in sagittal and transverse sections of the same brain, with maximal red depicting 8 percent shrinkage along a single scale dimension. Delineation of cortical and subcortical structures (lines in Figure \ref{fig:jacobian-mean-exvivo-invivo}) from the atlas mapping shows that the distortions are not uniformly distributed across the cortex. Thus, in future work quantifying cell or process densities, it would be important to take into account these local scale changes, rather than relying on an overall scale factor which may or may not reflect what is going on at a specific location in the brain. Note the importance of acquiring an in-vivo and an ex-vivo MRI scan, without which this analysis is not possible. 

The ex-vivo to histology coordinate mapping captures the deformative effects of the "sectioning" process. This mapping enables not only the histological reconstruction but also the segmentation of brain regions which allows the quantitative measurements of scale factor change.
In an analysis of the dataset of 15 brains, the cerebral cortical areas, thalamus, brainstem and the cranial nerves showed a local scale change of $< 1 \%$ per axis as measured by the percent scale change factor. The hippocampus, basal ganglia, pretectum and cerebellum showed a local change of 2-3$\%$ and the hypothalamus showed a local change of $ 4\%$ per axis. The in-vivo to ex-vivo coordinate mapping encompassing the extraction, perfusion, and fixation procedures shows much higher levels of distortions when averaged across the dataset. The cerebral cortical areas, thalamus, brain stem and the cranial nerves show a significantly larger absolute scale change of 5-6$\%$ per axis.
For the hippocampus, basal ganglia, pretectum, cerebellum, the difference was 6-8$\%$ for the in-vivo to ex-vivo maps. The hypothalamus showed a large change of $10\%$ per axis. 

\begin{figure}[H]
\vspace{-0.3cm}
\hspace{0.8cm}
\includegraphics[width=0.8\textwidth]{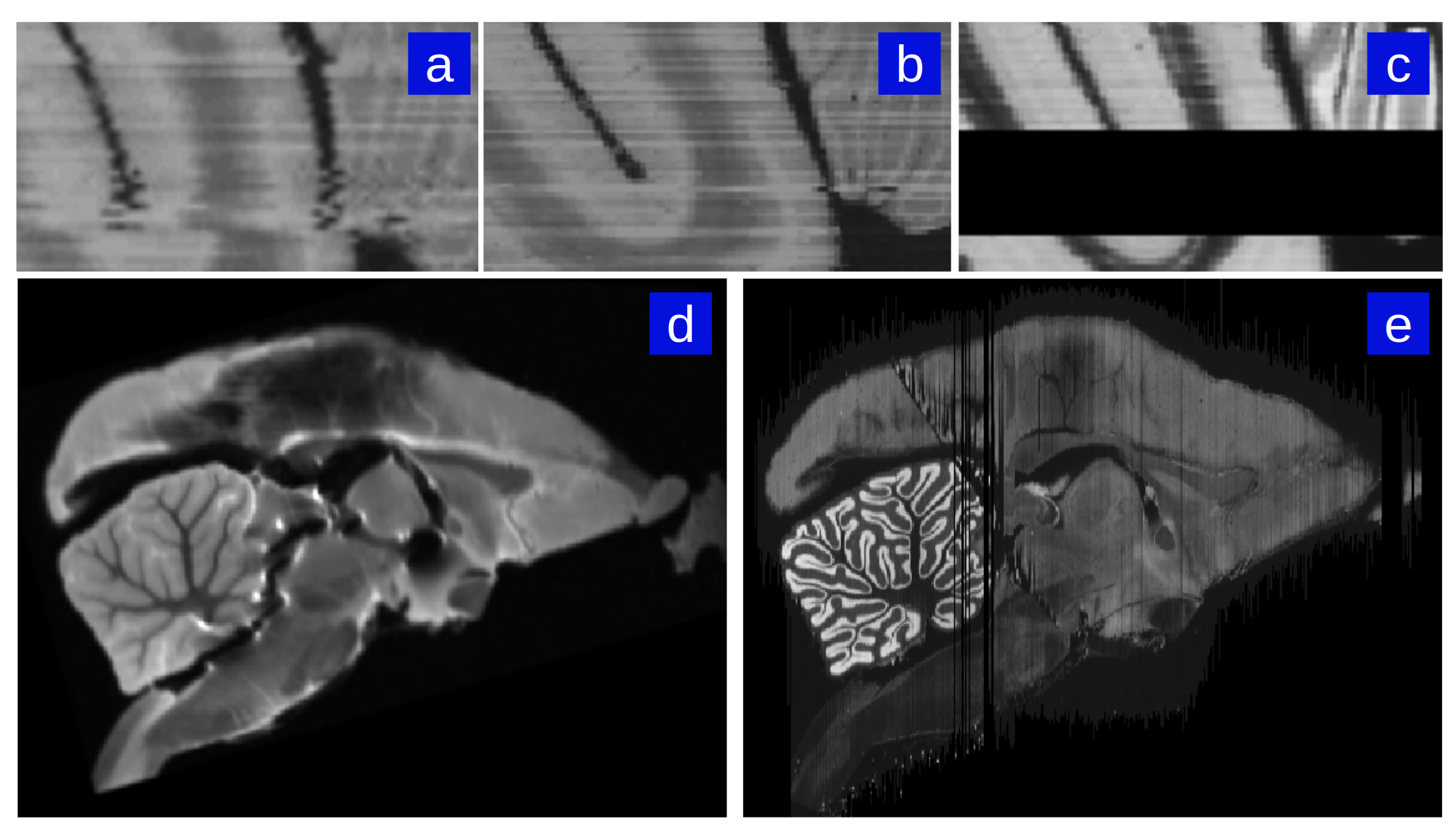} \\
\vspace{-0.7cm}
\caption{Reconstruction examples depicting the effect of the smoothness prior. When the Nissl stack (c) has missing or noisy sections, reconstruction to the next-best Nissl section contains small distortion accumulations (a). Inclusion of a smoothness constraint corrects this error (b). Additionally, highly damaged Nissl stacks can still be reconstructed despite major differences between damaged Nissl sections to corresponding MRI sections. For instance, ex-vivo MRI of a particular subject marmoset brain prior to sectioning (d) and successful Nissl reconstruction of the same marmoset brain despite major damage caused by sectioning process (e).}
\label{fig:smoothness}
\end{figure}

The constraint of anatomical smoothness within a brain volume is critical to producing accurate reconstructions, particularly in cases with {\it missing data or tissue damage} (for instance, the cross-modality registration presented here in the case where the Nissl stack is damaged or has missing sections). As part of our registration and reconstruction model described in Methods section 4.2, we previously introduced a smoothness prior in the form of a Sobolev derivative norm to provide robustness and control the dimension during the diffeomorphic mapping and restacking solution. The Sobolev prior couples adjacent sections and results in continuity of the reconstruction. It is particularly noticeable in the registration of multiple subject modalities to one another where sections are missing or damaged. The importance of this prior is visible in panels A-C of Figure \ref{fig:smoothness}.

The effect of the image intensity smoothness prior also manifests during Nissl-to-MRI stack reconstruction when there is significant damage to the Nissl brain. An example of our framework's ability to achieve accurate reconstruction when there is significant tissue damage (such as cutting and folding in particular sections) to the sections is shown in panels D and E of Figure \ref{fig:smoothness}.

\begin{figure}[H]
\begin{center}
\includegraphics[width=1.0\textwidth]{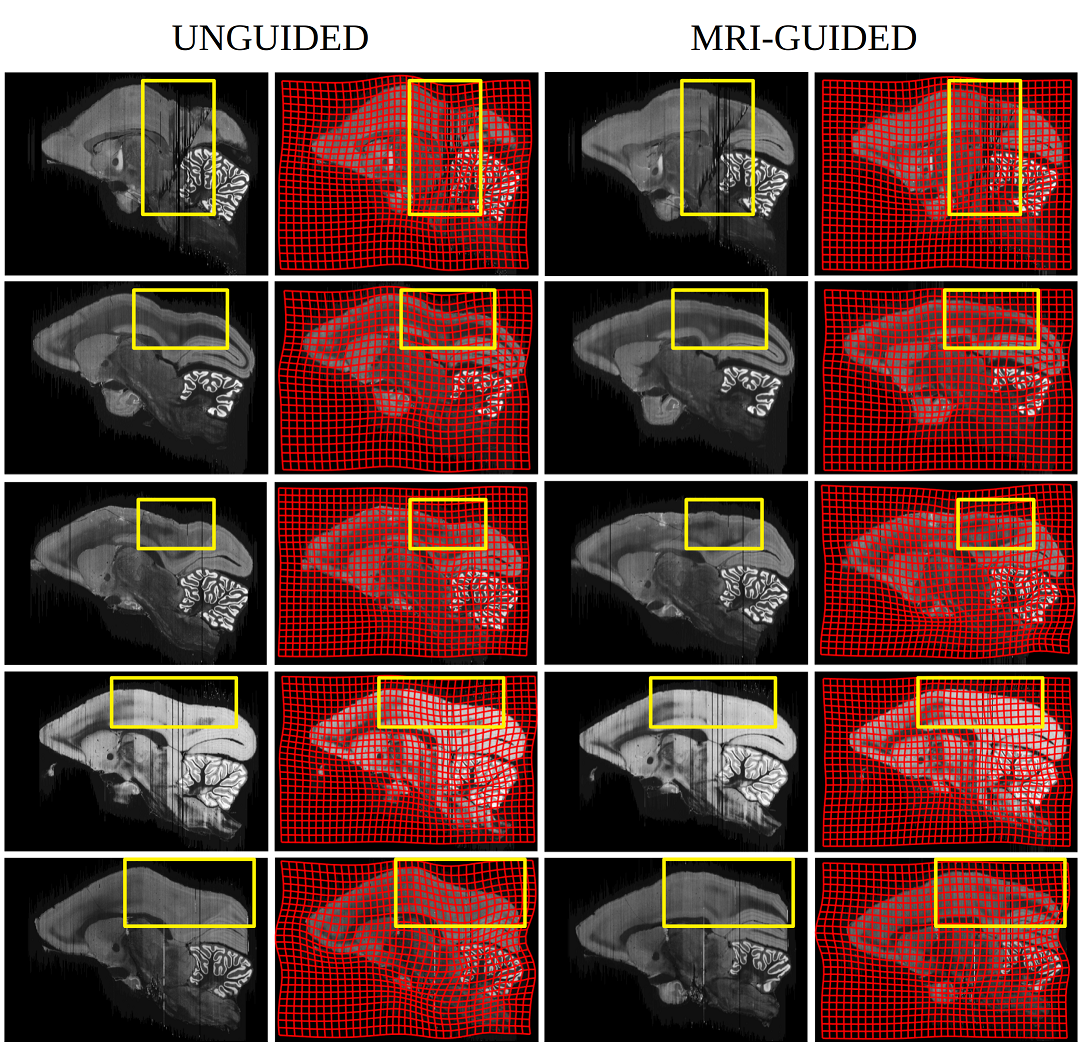} 
\caption{Histology restacking examples from Brain/MINDS dataset. Columns 1 and 2 depict sections 
exhibiting Malandain curvature artifact (notice cortical bend) associated to the unguided alignment with column 2 showing deformation of the underlying coordinate associated to the atlas-to-target warping; columns 3 and 4 show the Nissl histological stackings which are guided via the template based deformation.
Regions of large artifact compared to none are highlighted via yellow boxes.
\label{fig:malandain-many-sections}
}
\end{center}
\end{figure}

Our pipeline also allows for the correction of curvature artifacts associated with the histological restacking unguided by same-subject MRIs.
Figure 
\ref{fig:malandain-many-sections} shows examples of the curvature artifacts associated with 2D-3D reconstructions unguided by a reference brain \cite{Malandain2004}.
These figures demonstrate that the MRI guided registration pipeline solves the curvature issue. An example of this problem is shown in Figure \ref{fig:malandain-many-sections}. The left 2 columns
show the unguided stacking alignment which results in a large curvature artifact (highlighted within the yellow bounding box). The third and fourth columns show the guided restacking using our current pipeline. The yellow bounding boxes depict the areas with highest curvature bias effects. The curvature of the coordinate grid depicted in Figure \ref{fig:malandain-many-sections} is encoded by the 2D component of the $3 \times 3$ Jacobian matrix of the transformation.
Notice the curvature of the grid is more extreme for the unguided reconstruction which has no MRI to guide it globally.
The Malandain curvature artifact is present in the warping of the grids in the uncorrected restacking case due to the higher metric cost of $\varphi$ required to map an atlas onto the accumulated distortions of an unguided reconstruction.

\begin{figure}[H]
\includegraphics[width=1.0\textwidth]{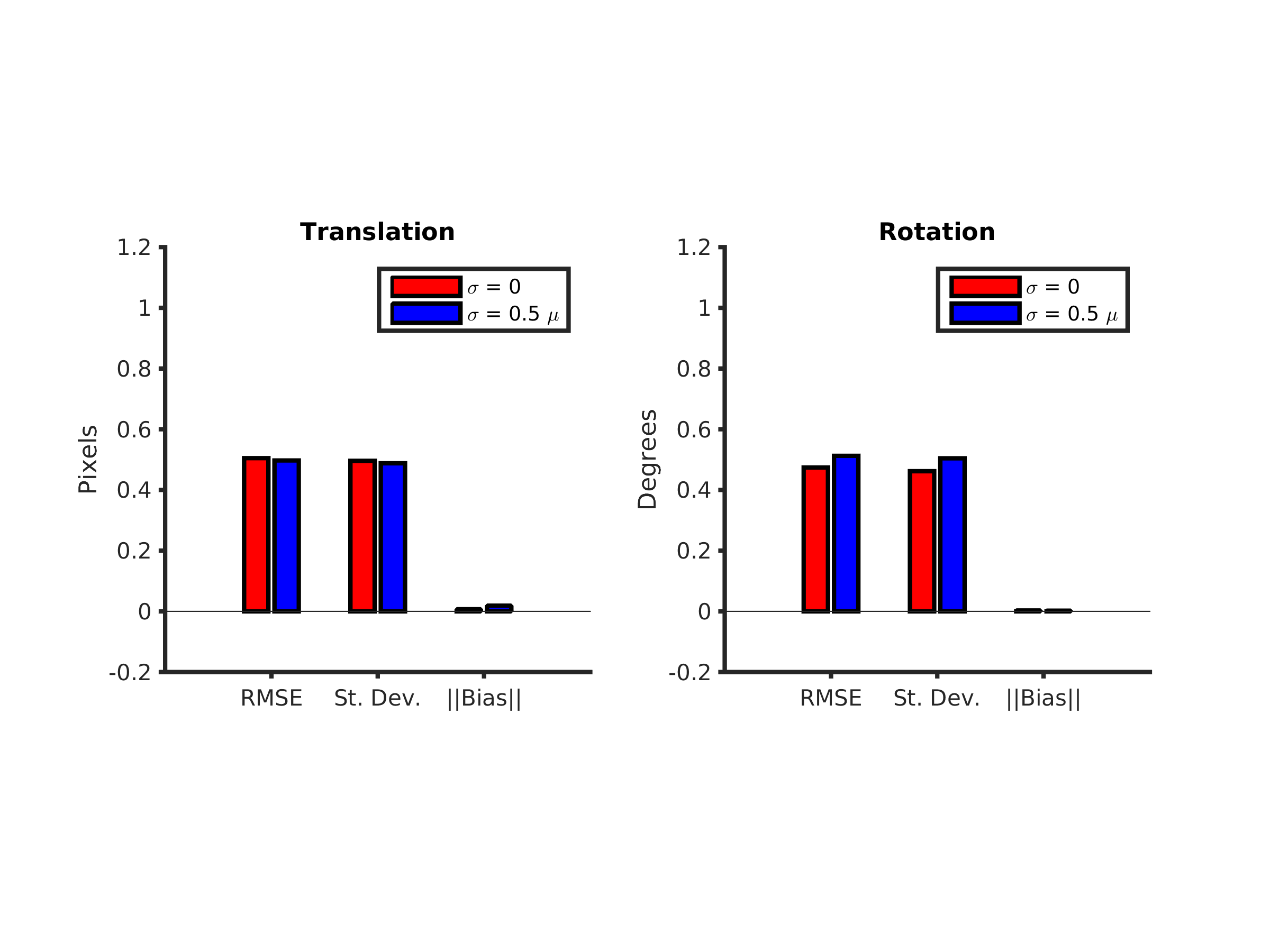} \\
\caption{RMSE, standard deviation, and norm of bias statistics for estimated MRI-guided reconstruction parameters from simulated Brain/MINDS phantom.}
\label{fig:simulation}
\end{figure}

The proposed method was also evaluated on a new simulated dataset (1000 examples) where sections of the Brain/MINDS atlas were randomly rotated and translated. The simulations demonstrated virtually unbiased estimates of the reconstructed stack with subvoxel standard deviations for both the translation and rotational aspects of the rigid motion. A summary of the simulation statistics is shown in Figure \ref{fig:simulation}. 
Included in Figure \ref{fig:malandain-many-sections} are clear illustrations of the use of the MRI atlas for dealing with the Malandain curvature artifact.

\section{DISCUSSION}
In this paper, we have shown the application of an LDDMM based registration pipeline to a multimodal dataset of marmoset brains. The methodology described in this paper is a significant advance over the unbiased re-stacking and integration to a reference atlas for a similar histological pipeline for the mouse brain, which did not include same-subject MRI volumes \cite{Lee-Mitra-Miller-PLOS-2018}.




Our results (Figure \ref{fig:jacobian}) reveal that the general deformation effect caused by the histological process is shrinkage in certain areas of the brain, and also expansion in other regions. The shrinkage is not surprising as it is generally well-known that some tissue shrinkage is caused by the histology procedures \cite{Mouritzen-1979}. Examination of the mean image (Figure \ref{fig:jacobian-mean-exvivo-invivo}) shows that shrinkage is not uniform throughout the brain but is generally located in the central and inferior regions of the brain, and near the ventricles. However, we note that some areas of the brain also showed expansions as depicted in the almost symmetric histogram of scale changes. More importantly, the methodology provides a {\it quantitative} measure for every brain voxel of the associated scale factor.

The diffeomorphic maps were generated for the population of 15 brains. The cubed-root of the determinant of the Jacobian was averaged across the empirical sample and the percent difference away from the identity was calculated for all of the structures within the template.
The average metric scale was computed for select brain areas such as the olfactory bulb, cortical areas, septum, thalamus, epithalamus, pretectum, brainstem, hypothalamus, cerebellum, cranial nerves, entorhinal cortex.
Our 3D volume reassembled maps from the tape transfer assisted histological sections matched very closely with the ex-vivo MRI maps. When the reassembled volumes from sections using the tape transfer technique \cite{mitra-tape-transfer-2015} were compared with the ex-vivo post perfusion MRI, the efficiency of the technique in preserving the tissue becomes evident.  

In contrast, quantification of the impact of the preparatory processes which was achieved by mapping the in-vivo MRIs to the ex-vivo MRIs, confirms the large, uni-directional shrinkage of brain tissue that has been reported in the literature. Again, we believe that this is the first time there is a detailed quantification of these changes brain-wide. We show that this shrinkage is not uniform across the brain and different brain areas show quite different levels of change. 

As first described by Malandain in the context of brain histology \cite{Malandain2004}, correct reconstruction of a sectioned object without prior knowledge of the object's shape is difficult and easily prone to error. Here, we presented models to inform our histological reconstruction algorithm with the best available shape prior -- either an MRI of the brain prior to sectioning or an atlas volume of the same species and modality. In the case of the reconstructions on the Brain/MINDS dataset, we showed results of the former model. For datasets that do not include same-subject reference imagery, the availability of whole brain 3D marmoset atlases, e.g. Paxinos \cite{Paxinos-marmoset-atlas} or Brain/MINDS \cite{Woodward-2018}, allow us to accurately reconstruct both the local and global restacking properties that are also free from the \cite{Malandain2004} curvature artifacts. 
Our simulation results show that same-subject reference guided reconstruction was able to recover the re-sectioning parameters with high accuracy, which not only provides value in improved visualization but also improves registration and segmentation accuracy.

When the shape prior is a same subject reference volume, the guided reconstruction acts as an improved initialization for nonlinear image registration, placing the voxels of the subject volume closer to their corresponding voxels in the atlas volume. As with any gradient-based optimization framework, LDDMM benefits from improved initialization as this reduces the likelihood of falling into a local minimum in the objective function. As shown in Figure \ref{fig:malandain-many-sections} the sample maps generated by registration of the Brain/MINDS atlas to an unguided reconstruction versus a guided reconstruction reveals an increased curvature of the underlying coordinate grid warp associated with the atlas mapping in the unguided case, indicating a displacement field with higher magnitudes and less homogeneity.

The addition of the smoothness prior via the Sobolev norm is valuable for providing robustness in the presence of noise or missing data or when the shape prior is not an exact reference volume. The driving intuition behind the smoothness prior is that in addition to the subject brain taking the shape of the reference volume, its image should be continuous and smooth. The effect of this prior is particularly noticeable in the registration of multiple subject modalities to one another where sections are missing or damaged. This is visible in the top row of Figure \ref{fig:smoothness}.

In conclusion, we have measured the local scale changes associated with the two major steps of histological processing for brain perfusion and sectioning reassembly. 
We measured the metric scale change using the fundamental form of the diffeomorphic mappings involved. We found that freezing and tape-transfer assisted histological sectioning produces minimallocal scale changes of $\sim 2\pm 0.4 \%$ per axis dimension. The preparatory step as assessed by mapping in-vivo MRI to ex-vivo MRI shows a much larger scale change of $\sim$ 6-8 $\%$ per axis dimension. To the best of our knowledge, this is the first systematic quantification of the local metric distortions associated with whole-brain processing.

\section{METHODS}

\subsection{Data}
The marmoset brain dataset used in this paper was prepared using a high throughput histological and image processing pipeline, described in \cite{Lin-2019-elife}. 
Briefly, each individual marmoset brain dataset consists of high resolution (9.4T) MRI scans (both in-vivo and ex-vivo post perfusion/extraction/fixation preparations) and high resolution images of a series of brain sections that have undergone histological processing to stain for Nissl substance, Myelin, Fluoroscent neuronal tracers and the expression of Cholera toxin B (CTB).  
The in-vivo MRI was acquired from the marmosets prior to any experiments. The ex-vivo MRI was acquired after the injection of tracers, the incubation period, perfusion, and fixation. The histological sections were imaged after sucrose cryoprotection, freezing, and cryo-sectioning. For simplicity, we refer to the procedures that occur between the in-vivo and ex-vivo MRI acquisitions as the "preparatory processes" and the procedures that occur between ex-vivo MRI acquisition and histology as the "sectioning process".
The histological image data was originally acquired at 0.46 $\mu$m in-plane resolution with 20 $\mu$m section thickness (alternating through four stains resulting in 80 $\mu$m gap between sections of a single modality, resulting in a $\sim$174x factor of anisotropy between XY and Z directions) and was downsampled to 80 $\mu$m in-plane resolution for computational purposes.

\begin{figure}[H]
\includegraphics[width=1.0\textwidth]{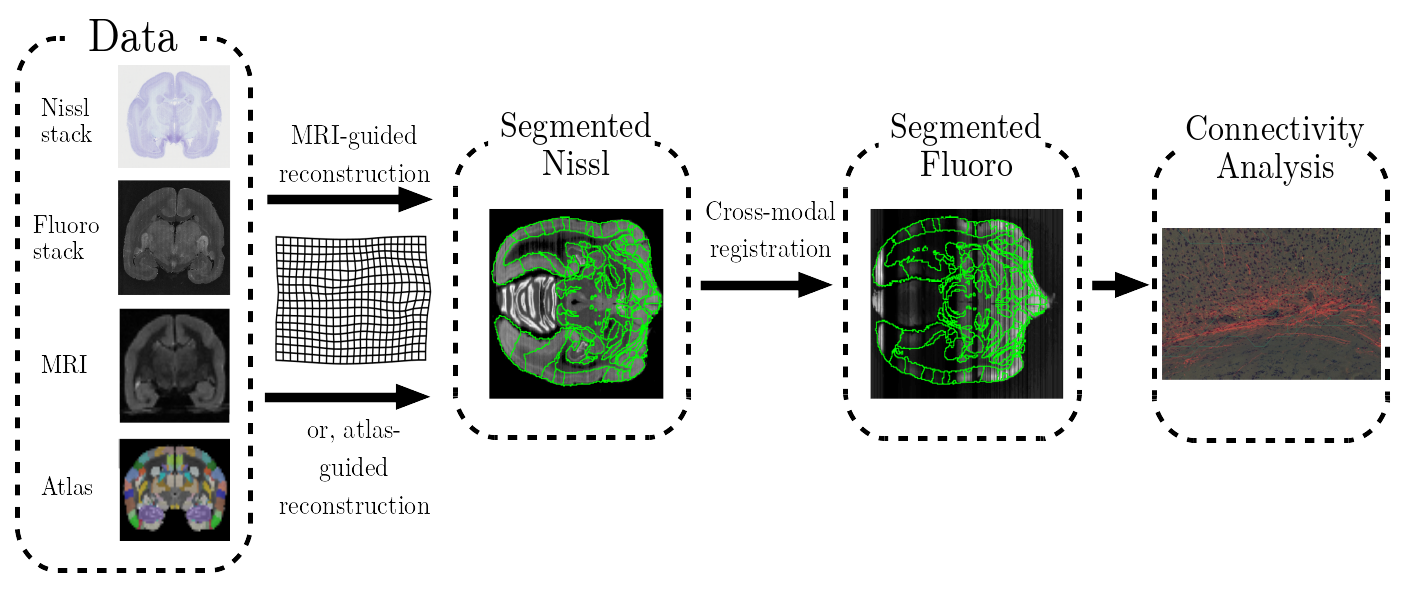} \\
\caption{Reconstruction pipeline workflow from multi-modality histological image sections to segmented data and connectivity analysis.}
\label{fig:pipelineFigure}
\end{figure}

In following sections, we describe two image registration algorithms (see Methods section 5.1) applied towards mapping the multimodal datasets of the marmoset brain (n=15) \cite{Lin-2019-elife} between the in-vivo MRI, ex-vivo MRI, and post-sectioning states within a single animal, as well as to a common population atlas. Firstly, we generated informed estimates of reconstructed stacks of Nissl-stained sections by simultaneously calculating a diffeomorphism $\varphi$ (a nonlinear smooth, 1-to-1, invertible mapping) from any given reference brain to the target coordinates and the unknown rigid jitter motions associated to each section (\textit{Model 2} in 5.1). Where available for a particular animal, ex-vivo MRI was substituted as the reference brain and is then matched to the histology stack using a low-dimensional linear rigid-scale/affine motion $ \varphi $ (6 or 12 degrees of freedom) of the MRI volume coupled to the rigid motions of the individual sections.
Gradient descent based algorithms were then implemented as described in \cite{Lee-Mitra-Miller-PLOS-2018}. A variant of the large deformation diffeomorphic metric mapping (LDDMM) algorithm \cite{Beg2005} that accounts for missing structures or damaged sections was also employed to compute maps between the Nissl based reference atlases and the reconstructed Nissl stained brain data \cite{Lee-Mitra-Miller-PLOS-2018}. For cross-modal registration of histological sections, we use a mutual-information based cross-modality matching function \cite{Kutten-Charon-2016}. The full computational approach including the multiple histological series is illustrated in Figure \ref{fig:pipelineFigure} in which both MRI as well as Nissl and fluorescent sections are shown.

\subsection{Models used in the pipeline: MRI to MRI model and Tape-transfer histology sectioning model }
The algorithmic framework proposed in this paper is based on the random orbit model \cite{GrenanderMiller2007}. 

In this paper, we explicitly examine two models: \textit{ Model 1} expresses cross-sectional anatomical variations solely via a diffeomorphism $\varphi$, assuming that the metric structure of the measured image $J$ is maintained as in MRI with only additive noise being added by the procedure. Here, $I$ is some exemplar deformed by $\varphi$. 
\textit{Model 2} includes the jitter caused by the histological sectioning and stacking, adding 3 degrees of freedom, $R_i$ for each z-section $i$.
The statistical models for each are defined as follows:
\begin{equation}
\nonumber
\begin{aligned}
&\text{Model 1: Volume-Volume} \ \ \
J(x,y,z) = I\circ \varphi^{-1}(x,y,z) + noise(x,y,z) \  ; \ \ \ \ \ 
\\
&\text{Model 2: Stack-Volume}\ \ \
J (R_i(x,y),z_i) =I \circ \varphi^{-1}(x,y,z_i) +noise (x,y,z_i) \ . 
\end{aligned}
\end{equation}

Estimates of the reconstructed stack $J^R = J(R_i(x,y),z_i), i=1,\dots,n$ are generated by simultaneously calculating the transform $\varphi$ from the atlas to the target coordinates and the unknown rigid jitter motions $R_i,i=1,\dots, n$ forming the iteratively reconstructed stack $J^R$. When $I$ is a population atlas image from a different subject, $\varphi$ is a diffeomorphism. When $I$ is a same-subject reference image like an MRI volume, $\varphi$ can be constrained to a rigid or affine transformation.

\subsection{Distances for Variational Methods}

The variational methods require building distances between the the mean fields in Models \textit{1} and \textit{2} and the histology stacks.

To build correspondences between the histological stack 2D sections $J^R(\cdot,z_i)$ and sections of the Nissl atlas or cross-modality MRI image $I(\cdot,z_i)$ we define a similarity metric based on either squared-error (same modality) or mutual information (cross modality) as described by Kutten et al \cite{Kutten-Charon-2016}.

\textbf{Squared-error within modality }
Define the error function between images $d:(I,J)\rightarrow R^+$ a positive squared-error function between images can be defined as the square of the Euclidean distance;
$$ d( I,J) = \frac{1}{2} \|J(\cdot) -I(\cdot)\|_2^2 = \frac{1}{2} \sum_{x,y.z} |I(x,y,z)-J(x,y,z)|^2
 \ .
$$
\textbf{Mutual information}
Across modalities,
$p_{I,J}$ is the empirical estimate of the joint histogram density and $p_I,p_J$ are the corresponding marginals.
The mutual information $d(I,J)$ is given by
\begin{equation}
d(I,J) = -\sum_\eta \sum_\psi p_{I,J}(\eta,\psi) \log \left( \frac{p_{I,J}(\eta,\psi)}{p_I (\eta) p_J (\psi)} \right) \ .
\end{equation}

\textbf{Decreasing the degrees of freedom for rigid motions}
Following \cite{Lee-Mitra-Miller-PLOS-2018} for the stacking reconstructions $J^R(\cdot,z_i)=J(R_i(x,y),z_i ), i=1,\dots,n$ we introduce a Sobolev smoothness penalty imposing a prior on the smoothness under the assumption anatomical structures are smooth arising from a smooth Sobolev space $H^k$ equipped with norm:
\begin{equation}
||J||^2_{H^k} = \sum_k \sum_{x,y,z} | \partial_h J(x,y,z)|^2  \ ,
\end{equation}
with $\partial_h$ denoting a total of $h$-derivatives in the three $x,y,z$ coordinates.

\subsection{Diffeomorphic atlas-mapping}
For generating diffeomorphic maps between atlas $\varphi: I \rightarrow J$ 
we use the LDDMM algorithm \cite{Beg2005} to generate flows $\dot \phi_t = v_t \circ \phi_t $, with $\varphi= \phi_1$.
The shortest flows that minimize the distance are calculated by defining an energy and calculating the variational minimizers of the flows:
\begin{equation}
\int_0^1 ||v_t||^2_V dt  + d(I \circ \phi_1^{-1}, J^R ) \ .
\end{equation}

For the $L2$ error we employ a multi-channel version of the LDDMM algorithm
\cite{CeritogluMultiModal2009}, defining multiple distances $|| C_n (I_n \circ \varphi^{-1}_1 - J_n^R ) ||_2^2$ associated to different masks.
Here, $C_n$ is a matrix of the same size and dimension as $I$ and $J^R$ taking on values from 0 to 1, where 1 represents full confidence in the subject data, 0 represents no confidence in the subject data (i.e. histological sections that are known to be excluded), and values in between indicate varying noise levels (i.e. partially damaged histological sections). This cost mask is multiplied element-wise with the similarity metric.

In the algorithmic framework, based on the random orbit model \cite{GrenanderMiller2007}, we explicitly model both the cross-sectional anatomic variations via smooth deformations of templates, as well as the random jitter introduced by the  histological sectioning process.
The biological anatomic variations are modeled as a smooth change of coordinates via diffeomorphisms of some set of exemplars or templates,  so that $I = I_{temp} \circ \varphi$ for some $\varphi \in Diff$
the diffeomorphism group, with 3D images $I(x,y,z), (x,y,z) \in {\mathbb R}^3$. The technical variations from the histology pipeline were modeled using rigid motions of a series of 2D sections with $z$-axis section coordinates $z_i ,i=1,\dots,n$. The random jittering of planar sectioning is represented as a series of rigid planar transformations $
R(x,y) =( x \cos \theta + y \sin \theta +t_i^x, 
-x sin \theta +y  \cos \theta    + t^y) $ with three degrees of freedom for each plane. Measurement noise was modeled as a spatial white noise, through the usage of an $L_2$ norm between the data and the template, or different data sets.

\subsection{Diffeomorphometry of Histological Procedures}
The first fundamental form of the mapping computed by the above method determines how vectors are transformed under mapping between coordinate systems (Atlas vs MRI vs histological image coordinates) and is specified by the Jacobian matrix $( \partial_X \varphi )$.
\begin{equation}
\partial_X \varphi(x,y,z)= \left ( \begin{array} {ccc}
\frac{\partial \phi_1(x,y,z)}{\partial x} &
\frac{\partial \phi_2(x,y,z)}{\partial x}&
\frac{\partial \phi_3(x,y,z)}{\partial x}
\\
\frac{\partial \phi_1(x,y,z)}{\partial y} &
\frac{\partial \phi_2(x,y,z)}{\partial y}&
\frac{\partial \phi_3(x,y,z)}{\partial y}
\\
\frac{\partial \phi_1(x,y,z)}{\partial z} &
\frac{\partial \phi_2(x,y,z)}{\partial z}&
\frac{\partial \phi_3(x,y,z)}{\partial z}
\end{array}
\right) \ .
\label{jacobian-matrix}
\end{equation}
The determinant $|\det \partial_X \varphi|$ and its logarithm are fundamental measures of coordinate change
and in the comparative study of the in-vivo and ex-vivo MRI and histology, directly measures the amount of
metric distortion within the same animal, and the change in measure across coordinate systems of different animals. The non-rigid component of the mapping is isolated by performing affine registrations between each coordinate space (in-vivo MRI, ex-vivo MRI, and histology) as a pre-processing step to diffeomorphic registration. The scale change assocated with the determinant of the affine transform matrix is included in the reported percent scale change.

\section*{Acknowledgments}

This work was supported by the G. Harold and Leila Y. Mathers Foundation, the Crick-Clay Fellowship, the H.N. Mahabala Chair, National Science Foundation Eager award 1450957, the National Institutes of Health [grant number P41 EB015909]; the National Institute on Aging [grant number R01 AG048349],
the Computational Anatomy Science Gateway as part of the Extreme Science and Engineering Discovery Environment (XSEDE) with grant number ASC140026, NIH DA036400, as well as the Kavli Neuroscience Discovery Institute supported by the Kavli Foundation.

%

\clearpage

\bibliographystyle{abbrv}
\bibliography{LDDMM}
\clearpage

\subsection*{Competing Interests}
M.I.M. reports personal fees from AnatomyWorks, LLC, outside the submitted work and jointly owns AnatomyWorks. This arrangement is being managed by the Johns Hopkins University in accordance with its conflict of interest policies. M.I.M.'s relationship with AnatomyWorks is being handled under full disclosure by the Johns Hopkins University. 

\subsection*{Corresponding Author}
Brian C. Lee (blee105@jhu.edu)

\clearpage
\appendix

\end{document}